\documentstyle[preprint,aps,fleqn]{revtex} 

\newfont{\bi}{cmbxti10 scaled\magstep2}
\begin{document}
\draft

\title{Critical Behavior of Nuclear-Spin Diffusion in GaAs/AlGaAs
Heterostructures near Landau Level Filling $\nu$=1}
\author{V. Bayot$^1$, E. Grivei$^1$, J.-M. Beuken$^1$, S. Melinte $^1$ 
and M. Shayegan$^2$}
\address{$^1$Unit\'e de Physico-Chimie et de Physique des Mat\'eriaux,
Universit\'e Catholique de Louvain,\\
Place Croix du Sud 1, 1348 Louvain-la-Neuve, Belgium\\
$^2$Department of Electrical Engineering, Princeton University,
Princeton N.J. 08544}

\date{\today}
\maketitle

\begin{abstract}
Thermal measurements on a GaAs/AlGaAs heterostructure reveal that the 
state of the confined two-dimensional electrons dramatically affects 
the nuclear-spin diffusion near Landau level filling factor $\nu$=1.
The experiments provide quantitative evidence that the sharp peak in 
the temperature dependence of heat capacity near $\nu$=1 is due to an enhanced
nuclear-spin diffusion from the GaAs quantum wells into the AlGaAs
barriers. We discuss the physical origin of 
this enhancement in terms the possible Skyrme solid-liquid phase transition.
\end{abstract}



In the presence of a perpendicular magnetic field ($B$), two-dimensional
electron systems (2DESs) exhibit striking phenomena originating from the
Landau quantization and electron-electron interaction; examples are 
the integral and fractional quantum Hall effects~\cite{QHEReview}.
At Landau level filling factor $\nu$=1, theoretical work on 2DESs in 
GaAs/AlGaAs heterostructures has shown that electron spin textures 
known as {\it Skyrmions} are the lowest energy, charged excitations of the 
ferromagnetic ground state~\cite{SkyrmeTheory}.
Skyrmions, which result from the dominance
of the electron-electron exchange energy over the Zeeman energy, are 
also the lowest energy state for quasi-holes and quasi-electrons {\it near} 
$\nu$=1; the ground state of a 2DES near $\nu$=1 is in fact expected to be 
a {\it crystal} of Skyrmions~\cite{Brey95,Green96,Cote97}.

Recent experimental studies have revealed that Skyrmions play a major role 
in many properties of quantum Hall 
systems~\cite{Tycko95,Barrett95,SkyrmeExp,Bayot96}.
In particular, measurements on a GaAs/AlGaAs multiple-quantum-well (QW) 
heterostructure indicated an anomalous behavior for the 
low-temperature apparent-heat-capacity ($C$) near $\nu$=1: $C$ 
is enhanced by a factor of up to $\sim$$10^5$ with respect to its low-$B$ 
value~\cite{Bayot96}. For temperatures $T\gtrsim$~0.1K, the magnitude as 
well as $T$- and $B$-dependence of the data are consistent with the 
large $C$ being dominated by the Schottky nuclear heat capacity of Ga and As 
atoms in the QWs.
This observation implies a strong enhancement of the nuclear spin-lattice
relaxation rate near $\nu$=1, consistent with the results of nuclear 
magnetic resonance (NMR) experiments~\cite{Tycko95,Barrett95}
and recent calculations~\cite{Cote97} which attribute the enhancement to the 
presence of low-energy spin-flip excitations in the 2DES.
An even more striking feature of the $C$ vs $T$ data is a 
very sharp peak at a very low temperature $T_c(\nu)$,
suggestive of a phase transition in the 2DES~\cite{Bayot96}.
While recent estimates for the melting temperature of the Skyrme 
crystal~\cite{Green96,Cote97}, including its dependence on $\nu$,
suggest that the peak in $C$ could originate from a Skyrme solid-liquid
phase transition, the physical mechanism that affects the nuclear spin dynamics 
and gives rise to the anomalous $C$-peak is not known and is a matter of 
debate~\cite{Green96,Cote97,Bayot96}.

In this Letter we report new quasi-adiabatic thermal experiments
revealing that the mechanism responsible for
the peak in $C$ vs $T$ is a dramatic enhancement of nuclear spin diffusion
across the QW-barrier interface. 
While only the nuclear heat capacity of the QWs' atoms is observed
away from $T_c$, the nuclei in both the QWs and barriers contribute to the
measured heat capacity near $T_c$.
We discuss the physical origin of this phenomenon in terms of the possible
Skyrme solid-liquid phase transition.


The experiments were performed on the same multiple-QW heterostructure used 
in Ref.~\cite{Bayot96}. It consists of 100 GaAs QWs separated by $\rm 
Al_{0.3}Ga_{0.7}As$ barriers. The wells and barriers are 250 and 1850$\rm 
\AA$ thick, respectively, and the barriers are $\delta$ doped with 
donors (Si) near their centers. We used a $\rm 7\times7 mm^2$ piece 
of the wafer which was thinned to 65$\mu$m.
The lattice $T$ is probed by a carbon paint resistor deposited on the substrate 
side of the sample and electrically connected 
with four very low thermal conductance, 8$\mu$m-diameter, NbTi wires. 
This resistor was calibrated against a RuO$_{2}$ resistance thermometer 
at $B$=0. 
Since the nuclear spin temperature ($T_N$) may 
be inhomogeneous in the sample due to a variable nuclear spin-lattice 
relaxation rate~\cite{Tycko95}, $T$ strictly refers to the measured 
{\it lattice} temperature.
The sample is supported by two 60$\mu$m-diameter NbTi wire which serves 
as the main weak thermal link to the cold finger of the dilution 
refrigerator mixing chamber.
In the restricted $T$ and $\nu$-range investigated here, $C$ is very
large and the observed external time constant, $\tau_{ext}$, reaches 
extremely large values which exceed $10^4$ seconds. 
Therefore, in the time scale of our experiments, 
which is governed by the observed internal time constant 
$\tau_{int}\sim 10^3$ s (see below), 
the sample is in the quasi-adiabatic regime. 
$B$ is applied at an angle of $\theta = 30^{\circ}$
with respect to the 2DES plane in order to match the experimental
conditions of Ref.\cite{Bayot96}.

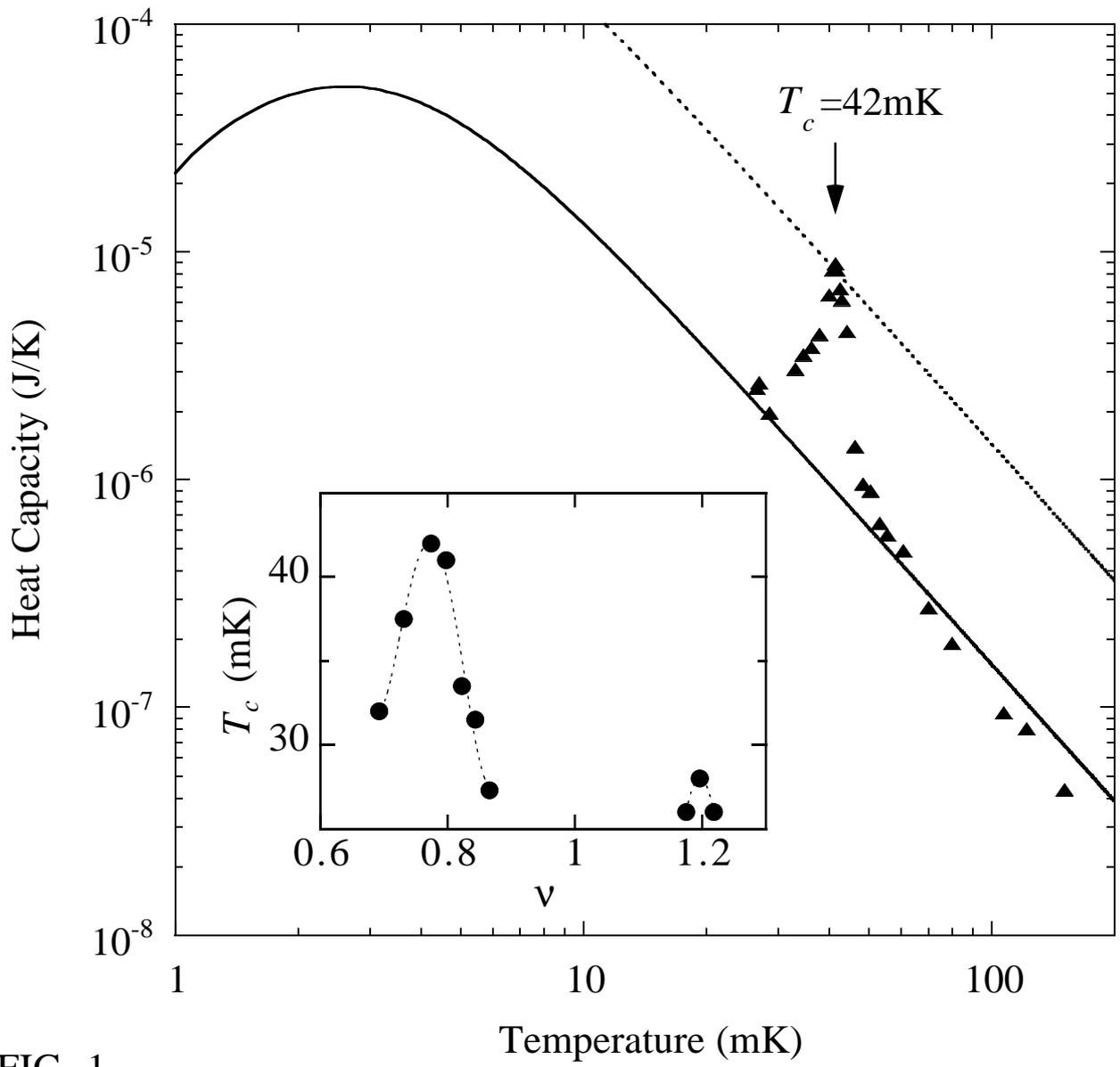
\begin{figure}
\caption{Measured $C$ vs $T$ at $B$=8.5T and $\theta = 30^{\circ}$ 
($\nu=0.77$). 
The curves represent the calculated Schottky nuclear heat capacity of the GaAs 
100-QWs (full curve) and the 100-period GaAs/Al$_{0.3}$Ga$_{0.7}$As 
heterostructure (dotted curve). 
The inset shows the measured $\nu$-dependence of $T_c$ at 
$\theta = 30^{\circ}$.}
\label{C_T}
\end{figure}

We first summarize the heat capacity results of Ref.~\cite{Bayot96}.
Figure~\ref{C_T} shows the measured $C$ vs $T$ at $\nu$=0.77, 
revealing a sharp peak at $T_c$=42mK. 
While at high $T$ ($\gtrsim$70mK) the measured value and $T$-dependence 
of $C$ are rather consistent with the calculated Schottky nuclear heat capacity 
of Ga and As atoms in the QWs~\cite{Bayot96,Schottky} (solid 
curve in Fig.~\ref{C_T}), at lower $T$, and especially near $T_c$, 
$C$ exceeds the calculated value by a factor of up to $\sim$10. 
The peak value of $C$, however, appears consistent with the Schottky 
nuclear heat capacity of the heterostructure if the nuclei of the 
{\it barrier} atoms are also included (dotted curve in 
Fig.~\ref{C_T}). This observation suggests that, near $T_c$, the 
barriers' nuclear spins also contribute to the measured $C$. 

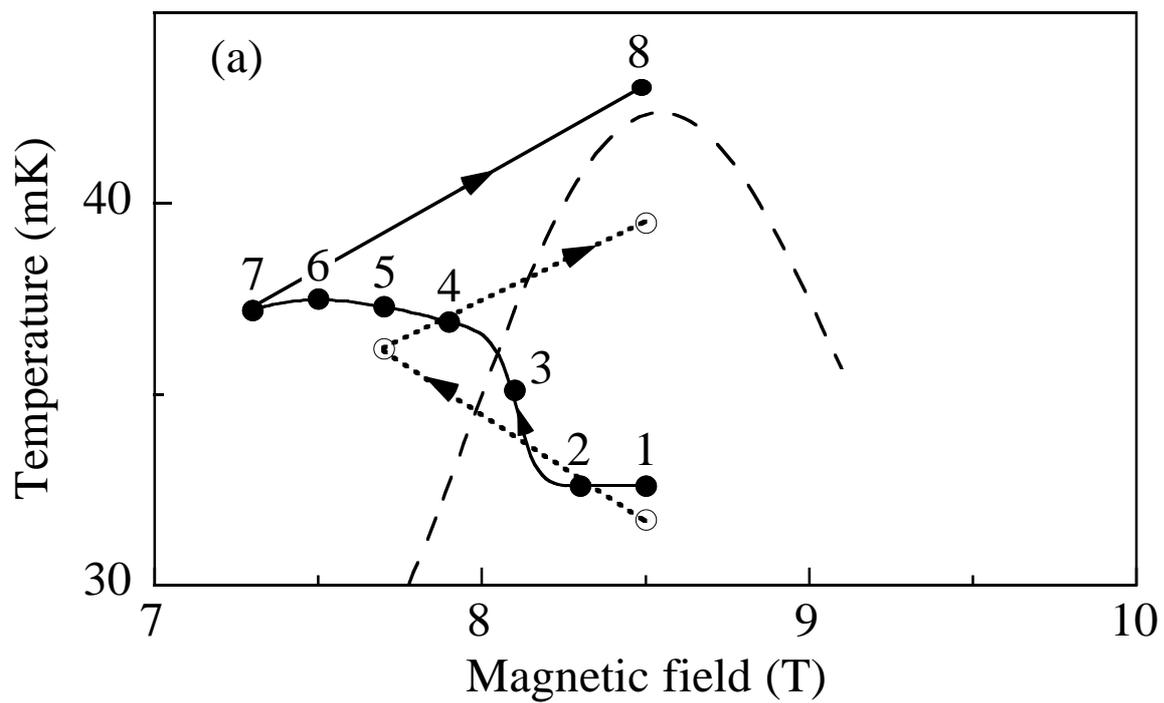
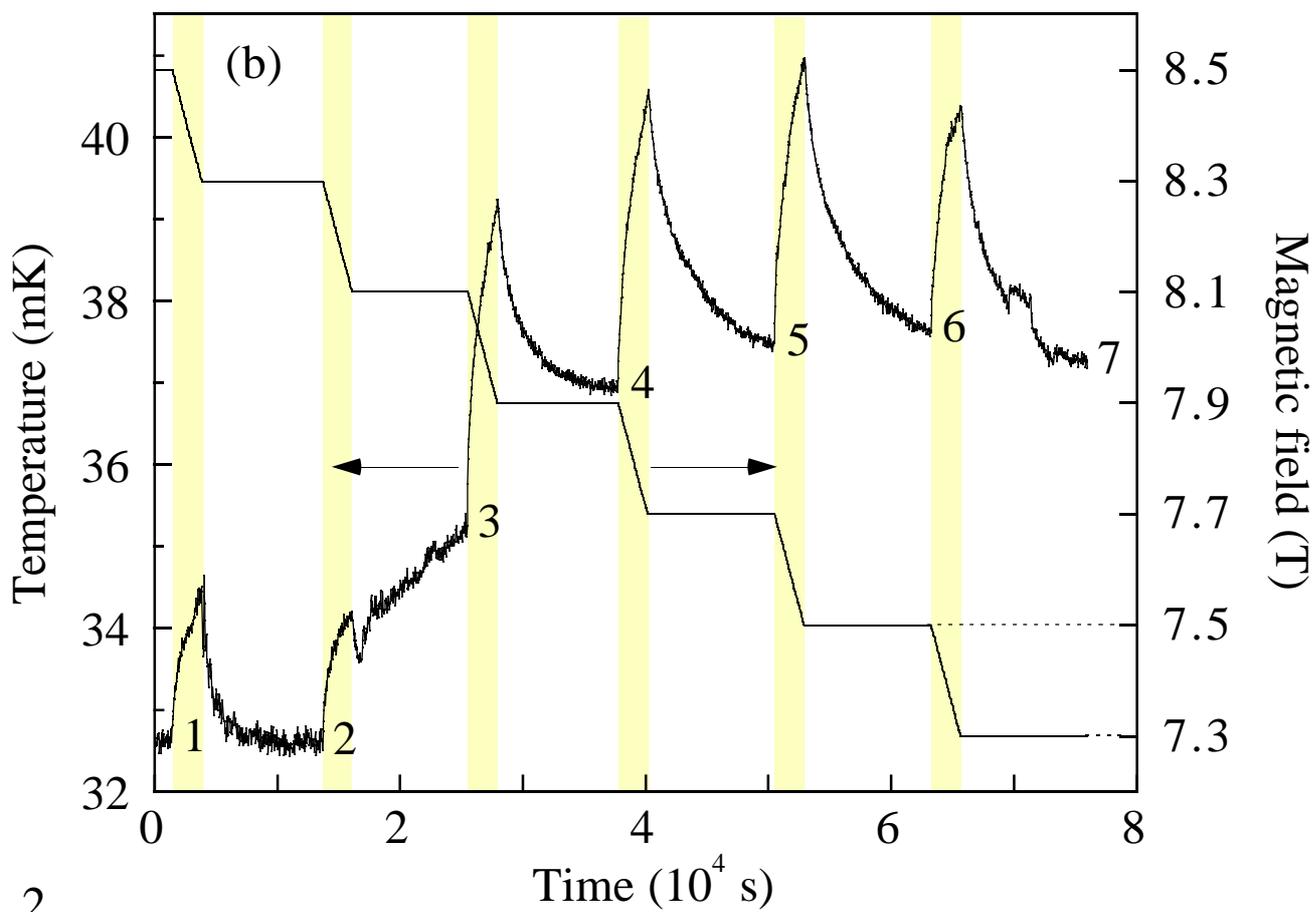
\begin{figure}
\caption{Measured lattice $T$ during the quasi-adiabatic experiments. 
(a) summarizes the evolution of $T$ during the 
first ($\circ$) and the second ($\bullet$) experiments (see text). 
The full and the dotted lines are guides to the eye and the arrows indicate the
direction of evolution. The dashed curve corresponds to the $B$-dependence of 
$T_c$ as reproduced from Fig. \protect \ref{C_T} inset. 
(b) shows $T$ and $B$ vs time in the second experiment. 
$B$-sweeps (gray areas) separate hold times
at the end of which $T$ is recorded (1 to 7) and summarized 
in (a) together with the final $T$ at $B$=8.5T (8).}
\label{Cycle}
\end{figure}

In our new quasi-adiabatic experiments, we utilize another noteworthy 
feature of the heat capacity data; as seen in Fig.~\ref{C_T} inset, $T_c$ 
sensitively depends on filling factor. This dependence, which is 
reproduced in Fig.~\ref{Cycle} (a) (dashed curve), is the key to 
understanding our results. In these experiments, we first fix the 
magnetic field at $B$=8.5T ($\nu$=0.77) where the peak in $C$ occurs at 
$T_c$=42mK (Fig.~\ref{C_T} and dashed curve in Fig.~\ref{Cycle}(a)). 
Starting from $T$$\sim$60mK, we cool down the cold finger to base 
temperature ($\sim$10mK) and the lattice slowly cools down
to $T$$\sim$32mK by waiting for about 50 hours.
Then, in order to further improve the adiabatic conditions, the weak heat flow 
that results from the $T$-difference between the lattice and 
the cold finger is reduced by increasing the cold finger 
temperature to 32$\pm$3mK.
Under these conditions, we observe that the measured sample $T$ is stable 
within the experimental accuracy ($\lesssim \pm$0.2mK) over periods longer
than $\tau_{ext}$, meaning that the sample is effectively quasi-adiabatic.

In a first experiment, whose results are shown by open circles in 
Fig.~\ref{Cycle}(a), we started from such a quasi-adiabatic condition 
and swept $B$ from 8.5 to 7.7T, at a rate of 0.01 T/minute. 
We observe that $T$ rises from 31.7mK to 36.2mK.  $B$ was then swept back 
to 8.5T at the same rate; this led to a rise of $T$ to 39.5mK. 
While the increase in $T$ with increasing $B$ may be explained by the 
adiabatic magnetization of the nuclei in the QWs, which couple to the 
lattice, according to \cite{Schottky}:
\begin{equation}
T_N\propto B,
\label{TN}
\end{equation}
the $T$ rise observed when $B$ is {\it lowered} from 8.5T to 7.7T is puzzling. 

In order to better understand the results of the above experiment, we performed
a similar experiment, starting from $T$=32.6mK, but here we swept $B$ in 
steps of 0.2T from 8.5T to 7.3T, with a hold time of 3 hours between steps. 
The evolution of $T$ during this second experiment is presented in 
Fig.~\ref{Cycle}(b) and summarized in Fig.~\ref{Cycle}(a) by close circles. 
We note that in each step, except for steps 2$\rightarrow$3 and 
3$\rightarrow$4, the lattice $T$ rises during the $B$-sweeps and then 
decays to approximately the same value as the one at the end of the 
previous step. 
These $T$ rises are due to Eddy-current heating of the 2DES and the lattice 
during the $B$-sweeps.  The heat is evidently slowly absorbed by the nuclei 
during and following the sweep via internal relaxation.  Because of the very
large heat capacity of the nuclear spins, this (Eddy-current) heating
does not lead to an appreciable increase in the lattice $T$ at the end
of the step \cite{Dev}. The key feature of this experiment, however, is
that during the 2$\rightarrow$3 and 3$\rightarrow$4 steps, the lattice $T$ 
increases significantly while the sample is in quasi-adiabatic conditions.  
Note in particular that, during the 2$\rightarrow$3 step, the lattice $T$ 
rises even {\it after} the $B$-sweep is completed. 
This striking behavior occurs when the lattice $T$ approaches $T_c$ at which
$C$ exhibits a maximum (dashed curve in Fig.~\ref{Cycle}(a)).  
Indeed, it appears that near $T_c$ {\it the lattice is heated internally}.

To explain the results of Figs.~\ref{C_T} and ~\ref{Cycle}, we propose a model 
in which the barriers' nuclei couple to the QWs' nuclei very near $T_c$, 
but decouple at higher and lower temperatures.  
The sharpness of the peak in $C$ vs $T$ data (Fig.~\ref{C_T}) 
is consistent with this model and implies that the coupling between the 
QWs and barriers is turned off rapidly as $T$ deviates from $T_c$.  
Therefore, we expect that during the initial cool-down at $B$=8.5T, when $T$ 
decreases below $T_c$=42mK, $T_N$ in the AlGaAs barriers remains close to 
42mK while $T_N$ of the QWs decreases down to $\approx$32mK due to strong 
coupling to the lattice. 
Next, decreasing $B$ below 8.5T  (steps 1$\rightarrow$7) in adiabatic 
conditions has two distinct consequences: (1) $T_N$ in both the QWs and
barriers is reduced due to adiabatic demagnetization of the
nuclei (Eq.~\ref{TN}), and (2) the coupling between the QWs and the barriers 
will increase dramatically when the dashed ($T_c$ vs $B$) curve in 
Fig.~\ref{Cycle}(a) is crossed.
While the former is essentially monotonic in $B$, the latter is not and 
implies that, near the dashed curve  in Fig.~\ref{Cycle}(a), 
$T_N$ should equalize over the whole GaAs/AlGaAs heterostructure. 
Since the barriers' nuclei were initially warmer than 
the QWs' nuclei, we expect a rise in lattice $T$ near this crossing, 
as observed experimentally. 
Finally, we attribute the $T$ rise during the final (7$\rightarrow$8) 
step to the nuclear magnetization of the nuclei (Eq.~\ref{TN}).

Beyond the good qualitative description of the experiments in
Fig.~\ref{Cycle}, the above model appears to provide a reasonable 
quantitative account of the data also.
First, when $B$ is lowered from 8.5T to 7.3T, we expect from Eq.~\ref{TN}
that nuclear demagnetization reduces $T_N$ in the barriers from 
$\approx$42mK to $\approx$36mK.
Since the barriers are by far thicker than the QWs, according to our model, 
this value should approximately correspond to the 
temperature of the entire heterostructure, and hence of the lattice. 
This prediction is in good agreement with experiments that give 
$T$$\approx$37mK at B=7.3T. 
The final increase of $T$ from $\approx$37mK to $\approx$43mK when 
$B$~increases from 7.3T to 8.5T is consistent with the nuclear 
magnetization of the heterostructure (Eq.~\ref{TN}).
Second, the fact that, at $T_c$, $C$ reaches the value expected for the entire 
heterostructure (dotted curve in Fig.~\ref{C_T}) is clearly consistent with 
our model and hence provides additional credence to our interpretation.

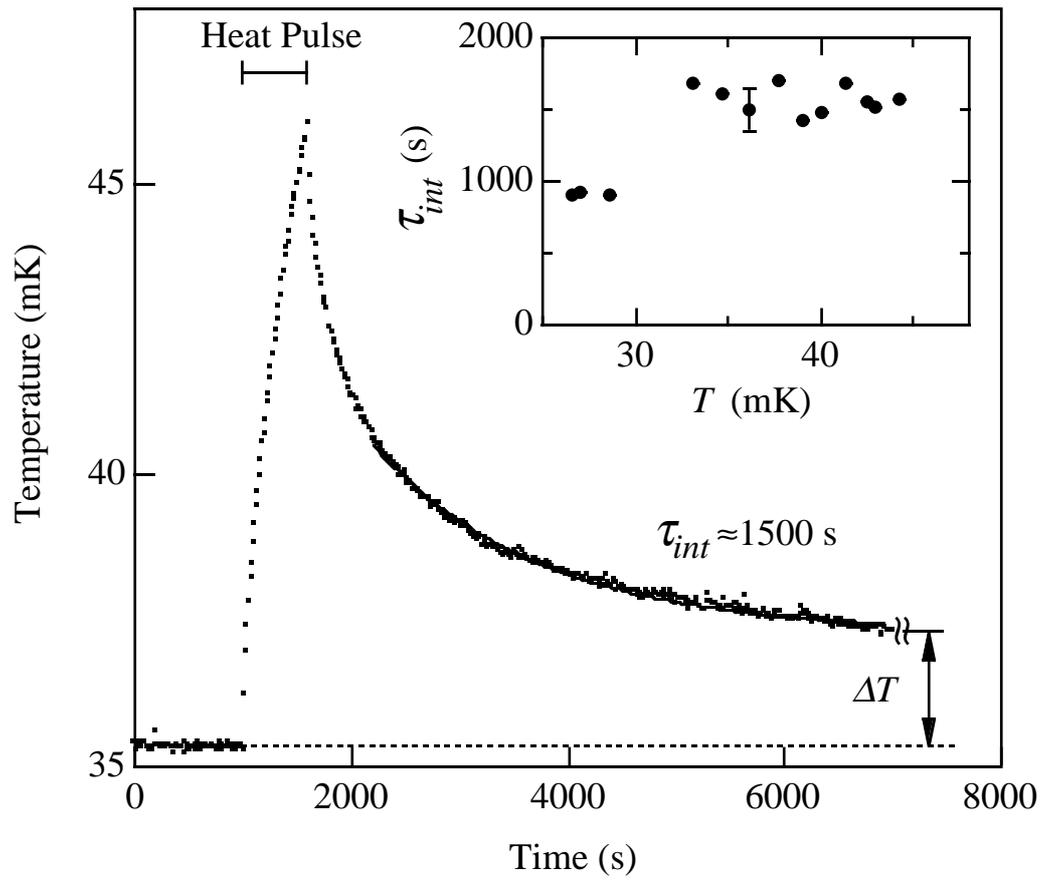
\begin{figure}
\caption{$T$ vs time during a heat capacity experiment in the quasi-adiabatic 
regime ($C = Q/\Delta T$ where $Q$ is the heat injected during the pulse; 
$B$=8.5T and $\theta = 30^{\circ}$). 
The heat pulse is followed by a relaxation characterized by an exponential
decay (full curve) giving $\tau_{int}\approx 1500$s. 
The inset shows $\tau_{int}$ vs $T$ at the same $B$ and $\theta$.}
\label{pulse}
\end{figure}

Next, we discuss the internal time constant ($\tau_{int}$) observed in our 
experiments. 
Figure~\ref{pulse} presents a typical heat pulse/relaxation trace, in 
quasi-adiabatic conditions, obtained during heat capacity experiments 
near $T_c$.
The relaxation follows an exponential decay characterized by $\tau_{int}$. 
The inset to Fig.~\ref{pulse} shows that $\tau_{int}$ exhibits two different, 
but remarkably constant, values above and below 30mK: $\sim$1600s and 
$\sim$900s, respectively. We note that this crossover temperature 
corresponds to the $T$ below which $C$ is reduced back to a value close to the 
nuclear heat capacity of the 100-QWs, $C_{QW}$ (full curve in Fig.~\ref{C_T}). 

In order to explain these observations, we consider three possible 
mechanisms responsible for the observed $\tau_{int}$. 
Heat diffusion in the nuclear spin system of the barriers 
is governed by the one-dimensional diffusion equation:
\begin{equation}
{\partial T \over \partial t}=D{{\partial }^{2}T \over {\partial z}^{2}}
\label{Diff}
\end{equation}
where $D\sim 10^{-17} \rm m^2/s$ is the nuclear-spin diffusion 
constant in GaAs \cite{Paget82} and the $z$-axis is along the growth direction. 
Therefore, it gives an internal time constant:
\begin{equation}
\tau_{d}\sim {r^2 \over D}
\label{tau}
\end{equation}
where $r$ is the distance over which diffusion takes place\cite{Tye}. 
Since each barrier is surrounded by two QWs, we take $r$=925$\rm \AA$ 
(half the barrier thickness) and Eq.~\ref{tau} gives $\tau_{d}$$\sim$800s.
The two other mechanisms involved in the  heat transfer from the lattice to the 
barrier nuclei, i.e. nuclear spin-lattice relaxation in the QWs and diffusion 
across the QW-barrier interfaces, result in two additional time constants: 
$T_{1}$ and $\tau_{i}$, respectively. 
When $C \approx C_{QW}$ ($T\lesssim 30$mK), the diffusion across the 
QW-barrier interface is negligible and $\tau_{int}$ should be determined by 
$T_{1}$ only. 
On the other hand, when $C > C_{QW}$ ($30\lesssim T\lesssim 50$mK)
at least a fraction of the barriers' nuclei couple to the QWs' nuclei. 
This implies that 
$\tau_{d}$ and $\tau_{i}$ should {\it increase} $\tau_{int}$. 
This is consistent with the rise  in $\tau_{int}$ observed above 30mK 
(Fig.~\ref{pulse} inset).
Moreover, we note that the increase of $\tau_{int}$ 
above 30mK ($\sim$700s) is comparable to the estimated 
$\tau_{d}$$\sim$800s, implying that $\tau_i$ is rather small once the 
barrier nuclei do couple to those in the QWs\cite{substrate}.

We now remark on the physical origin of the peak in $C$ vs $T$.
NMR experiments show that nuclear spin-diffusion from the QWs into
the barriers is extremely weak, except when optical pumping broadens the
Knight shift peak of the QWs which then overlaps with the NMR peak of the
barriers\cite{Tycko95}. The spectral overlap allows spin-diffusion
which is driven by nuclear magnetic dipole-dipole coupling\cite{Tycko95,DDC}.
Therefore, the enhancement of spin-diffusion across the QW-barrier interface 
near $T_c$ could originate from a broadening of the Knight shift peak in the QWs.
In the liquid Skyrme phase, motional averaging of the local 2DES spin 
polarization produces a single Knight shift peak \cite{Brey95,Barrett95}. 
On the other hand, the absence of motional averaging in the solid Skyrme 
phase should induce both positive and negative Knight shifts depending on the 
{\it local} spin polarization of the 2DES. Since above and below the 
peak in $C$ vs $T$ the diffusion across the QW-barrier interface is very weak, 
this implies that in both the liquid {\it and} the solid Skyrme states, 
the Knight shift peak(s) do not overlap with the NMR peak of the barrier.
One possible explanation for the $C$ anomaly is that the critical slowing down of the 
spin fluctuations in the 2DES,  
associated with the Skyrme liquid-solid phase transition\cite{Cote97}, 
could induce a broadening of the Knight shift peak in the QWs. 
This broadening would induce spectral overlap of the QWs' and barriers' 
NMR peaks and hence allow spin-diffusion across the QW-barrier interface 
{\it only} near $T_c$.

Finally, it is worth emphasizing that the shape of the peak in $C$ vs $T$
points to a second-order phase transition \cite{peak}. 
The fact that $\tau_{int}$ is constant in the $T$-range where 
$C > C_{QW}$ near $T_c$ implies that the {\it strength} of the coupling to the 
barriers remains constant while $C$ varies strongly. 
Therefore, it appears that the smaller value of $C$ on the  sides of the peak in 
$C$ vs $T$ comes from a {\it reduced fraction} of the sample in which the 
barriers' nuclei couple to the QWs' nuclei. 
The complexity of spin textures in the 2DES, together with the fact that
the phase transition is not observed directly but through its effect on 
nuclear-spin lattice relaxation, does not allow a quantitative analysis 
of the data in the framework of the existing theories that describe
order-disorder phase transitions. On the other hand, Skyrmions form an 
XY ferromagnet which has broken translation symmetry and one 
expects at least one phase transition as $T$ is lowered \cite{Cote97}.


In conclusion, our experiments reveal the critical influence of the
ground and excited states of the 2DES near $\nu$=1 on nuclear-spin dynamics 
in GaAs/AlGaAs heterostructures. 
Our results provide a quantitative phenomenological description of the 
apparent heat capacity anomaly that may originate from a Skyrme solid-liquid 
phase transition. 


The authors are much indebted to S.E. Barrett, S.M. Girvin and A.H. MacDonald
for fruitful discussions and suggestions.
This work has been supported by NATO grant CRG 950328 and the NSF
MRSEC grant DMR-9400362.
V.B. acknowledges financial support of the Belgian National
Fund for Scientific Research.



\begin{thebibliography}{10}
\bibitem{substrate} It is worth remarking that the parts of the sample for which
$\tau_{d}\gg \tau_{ext}$ do not contribute to the measured $C$ in the $T$ and 
$B$ range we are discussing. Since $\tau_{d}\gtrsim 10^5$ s for 
$r > 10^{-6}$ m 
(Eq.~\protect \ref{tau}), the GaAs substrate contributes negligibly to $C$.
\end{thebibliography}
\end{document}